# The “seesaw” impacts between reduced emissions and enhanced AOC on $O_3$ during the COVID-19

Shengqiang Zhu[1], James Poetzscher[1,2], Juanyong Shen[3], Siyu Wang[1], Peng Wang[4*], Hongliang Zhang[1,5*]

[1]Department of Environmental Science and Engineering, Fudan University, Shanghai 200438, China

[2] School of Environmental Science and Engineering, Nanjing University of Information Science and Technology, 219 Ningliu Road, Nanjing 210044, China

[3]School of Environmental Science and Engineering, Shanghai Jiao Tong University, Shanghai 200240, China

[4]Department of Civil and Environmental Engineering, Hong Kong Polytechnic University, Hong Kong 99907, China

[5]Institute of Eco-Chongming (IEC), Shanghai 200062, China

*Corresponding authors: Peng Wang, peng.ce.wang@polyu.edu.hk; Hongliang Zhang, zhanghl@fudan.edu.cn.

**Abstract**

Due to the lockdown measures during the 2019 novel coronavirus (COVID-19) pandemic, the economic activities and the associated emissions have significantly declined. This reduction in emissions has created a natural experiment to assess the impact of the emitted precursor control policy on ozone ($O_3$) pollution, which has become a public concern in China during the last decade. In this study, we utilized comprehensive satellite, ground-level observations, and source-oriented chemical transport modeling to investigate the $O_3$ variations during the COVID-19 in China. Here we found that the $O_3$ formation regime shifted from a VOC-limited regime to a $NO_x$-limited regime due to the lower $NO_x$ during the COVID-19 lockdown. However, instead of these changes of the $O_3$ formation region, the significant elevated $O_3$ in the North China Plain (40%) and Yangtze River Delta (35%) were mainly attributed to the enhanced atmospheric oxidant capacity (AOC) in these regions, which was different from previous studies. We suggest that future $O_3$ control policies should comprehensively consider the synergistic effects of $O_3$ formation regime and AOC on the $O_3$ elevation.

## Introduction

Beginning in January 2020, a novel coronavirus (COVID-19) began rapidly spreading throughout China, first in Wuhan, and then in other major cities [1, 2]. In response, to prevent the spread of COVID-19, the first lockdown was implemented on January 23rd, suspending non-essential traffic in Wuhan, the epicenter of the outbreak. Over the following days, major cities throughout China issued similar travel restrictions and stringent lockdown measures, affecting over half a billion residents. These restrictive measures resulted in substantial reductions in human activities, which consequently induced unprecedented decreases in anthropogenic emissions of air pollutants, especially from the industry and transportation sectors [3, 4]. In particular, during the lockdown, $NO_2$ levels in Eastern China were estimated to have decreased by ~65% compared to the same period in 2019, mainly due to the reduction of vehicle emissions[5, 6]. Similarly, other pollutants such as $SO_2$ and CO also decreased. In contrast, during the same period, ozone ($O_3$) increased significantly with a nationwide increase of 47.3% [5, 7, 8].

Besides the lockdown period, China experienced persistent $O_3$ pollutions in recent five years, especially in urban areas [9-11]. $O_3$ is formed through non-linear photochemical reactions of nitrogen oxides ($NO_x$ = NO + $NO_2$) and volatile organic compounds (VOCs) [12]. The $O_3$ sensitivity regime, determined by the relative abundance of VOCs and $NO_x$, plays a significant role in $O_3$ formation [13, 14]. Previous studies have reported that the elevated $O_3$ during lockdowns was mainly attributed to the enhanced atmospheric oxidation capacity (AOC) [15, 16], reflected by the levels of major oxidants such as hydroxyl radical (OH) and nitrate radical ($NO_3$) [17]. Specifically, $NO_2$ is defined as the main sink of OH radical through the reaction to form nitric acid ($HNO_3$) [18, 19]. During the lockdown, the drastic decreases in $NO_2$ levels increased OH concentration, which then reacted with VOCs, facilitating the formation of secondary pollutants [20]. Unfortunately, the current understanding of AOC and secondary pollution is still limited [21-23]. The COVID-19 lockdown provides an important opportunity to study the interaction between emissions, AOC, and meteorological conditions.

In this study, we used observational and satellite data to identify changes in $O_3$ levels and its associated precursors ($NO_2$ and HCHO) during the COVID-19 lockdowns in China. The Community Multiscale Air Quality (CMAQ) model was also applied to analyze the characteristics of air quality in the same period. The roles of the $O_3$ sensitivity regime and AOC were also discussed to provide an in-depth explanation for the increase in $O_3$. We found that $O_3$ elevations during the COVID-19 lockdown period in the NCP and YRD were mainly controlled by enhanced AOC, even suppressing the impacts of the $O_3$ formation regime due to the reductions of anthropogenic emissions. In contrast, $O_3$ decreased slightly in the PRD. The results aim to formulate more effective emission control policies, particularly focused on reducing AOC to battle the persistent $O_3$ pollution in China.

## Results

**Significant $O_3$ variations during the Lockdown.** Here we investigate $O_3$ changes during three periods around the COVID-19 Lockdown, which are defined as Pre-lockdown (January 6 to January 22, 2020), Lockdown (January 23 to February 29, 2020), and Post-lockdown (March 1 to March 31, 2020), respectively.

According to surface observation, changes in China's surface MDA8 $O_3$ show significant spatial variations from Pre-lockdown to Post-lockdown. During the Lockdown, elevated $O_3$ occurred in large areas throughout northern and central China, while it decreased in South China (Fig. 1a), consistent with previous studies [5, 15, 24]. The most prominent $O_3$ increase occurred in the NCP (Fig. 1b), with a mean MDA8 $O_3$ increase of 54% (from 24 ppb to 37 ppb) (Fig.1c). In Baoding and Shijiazhuang (major cities in the NCP), $O_3$ increased by over 100%. Moreover, in the YRD, a noticeable MDA8 $O_3$ increase of 44% (from 26 ppb to 38 ppb) was observed. During Post-lockdown, observed $O_3$ concentrations continued to increase in the NCP and YRD, partially due to the rising temperature (Fig. S2). $O_3$ variation is more complex in the PRD, however. In general, $O_3$ levels decrease from Pre-lockdown to Post-lockdown. But in Guangzhou, the most populated city of PRD, an increase in $O_3$ was observed. Considering the similar temperature levels between Pre-lockdown and Lockdown over China, these variations are more related to the sudden reductions of $O_3$ precursors.

In addition, heightened $O_3$ pollution was observed in the NCP and YRD during the Lockdown compared to the same period in 2019 (Fig. S3). Although $O_3$ precursors decreased drastically in these regions, mean MDA8 $O_3$ levels were 14-19% higher than in 2019. In contrast, in the PRD, the mean MDA8 $O_3$ during the Lockdown of 2019 is close to (or even slightly higher) that in 2020, and the opposite of the trend observed in the other regions. Similar temperature conditions are also found during the Lockdown in 2019 and 2020, which fails to fully elucidate the $O_3$ differences between these two years, demonstrating that the emission reduction of $O_3$ precursors plays a more critical role in $O_3$ variations. Considering the $O_3$ results in the Lockdown periods in 2019 and 2020, it is critical to deeply understand the seesaw phenomenon between elevated $O_3$ and its reduced precursors.

**Changes of $O_3$ precursors and formation regimes.** Given that the ratio of HCHO to $NO_2$ determines the $O_3$ formation regimes, HCHO and $NO_2$ are considered the most important precursors of $O_3$ [25]. The satellite column data and CMAQ model have revealed significant reductions of $NO_2$ throughout much of China, especially in NCP and YRD regions (Fig S4a). According to the satellite data, $NO_2$ in the NCP, YRD, and PRD regions declined by 59.61%, 63.28%, and 44.03% during the Lockdown respectively. These reductions are mainly attributed to the significant decline of $NO_x$ emissions from industry, power, and transportation sectors illustrated by the source apportionment analysis (Table. S6).

However, no noticeable changes were observed in the HCHO concentration during the Lockdown. The spatial distribution of HCHO, similar to that of $NO_2$, exhibits higher levels in southeast China, whereas in western China, due to the low anthropogenic VOCs emissions, the HCHO concentration is relatively low [26, 27]. The HCHO in the atmosphere is mainly formed through direct emissions from industrial and biogenic sectors and through secondary sources such as the oxidation reaction between VOCs and OH. During the Lockdown, emissions of HCHO and other VOCs declined significantly (-37%) in China (Table.S1) and therefore might have reduced HCHO levels. However, the enhanced AOC [5] during Lockdown likely promoted the formation of HCHO from secondary sources, offsetting the impact of the decline in HCHO emissions and explaining why HCHO levels remained relatively constant. Also, the oxidation of methane, which has a long lifetime and relatively stable concentrations, plays an important role in the HCHO concentration [28, 29], which also might help maintain constant HCHO levels as shown in our source apportionment analysis (Table.S6) during the Lockdown period.

In general, the $O_3$ sensitivity regimes in China shifted from VOC-limited to $NO_x$-limited during the Lockdown, as indicated by both satellite data and model simulations (Fig. 2b). During the Pre-Lockdown period in 2020, the VOC-limited regime dominates in the NCP, YRD, and PRD regions due to the relative abundant $NO_x$ emissions from industry and transportation sectors, consistent with the previous studies [30]. However, during the Lockdown period, VOC-limited regimes transitioned to $NO_x$-limited regimes in these regions. The percentage of $NO_x$-limited regimes in the NCP, YRD, and PRD regions during the Lockdown period increased from 11%, 37%, and 31% to 56%, 65%, and 69%, respectively. These changes in the $O_3$ formation regime and $NO_2$ and HCHO concentrations were not observed in 2019 (Fig. S5a); $NO_2$ and HCHO concentrations during the periods in 2019 that correspond to the Pre-lockdown and Lockdown periods in 2020 remained relatively constant, explaining the lack of variation in the $O_3$ formation regimes (Fig. S5b).

Surprisingly, $O_3$ levels increased in the $NO_x$-limited regimes in the NCP and YRD regions during the Lockdown period in 2020, even though pronounced reductions of $NO_x$ levels were observed in these regions. This unusual phenomenon is contrary to previous studies [28], indicating that the enhanced AOC might have played a more significant role in increasing $O_3$ in these regions during the Lockdown period in 2020.

**The dominating role of the enhanced AOC in $O_3$ formation.** Our model simulations demonstrated significantly enhanced AOC in the NCP and YRD regions, which is consistent with the variation of $O_3$ concentrations. $HO_x$ (OH and $HO_2$) radicals, the main daytime oxidant, increased significantly in central and northern China with the highest growth rates of 0.06 and 2.71 ppt for OH and $HO_2$ radical, respectively due to the relatively low levels of $NO_2$, the primary $HO_x$ sink (Fig. 3a and Fig. S5a). Specifically, in the NCP, YRD, and PRD regions, the average increase in $HO_x$ was 0.79, 0.92, and 0.17 ppt, respectively. The rise in OH and $HO_2$ radicals could be the leading cause of the $O_3$ increase during the Lockdown period in the NCP and YRD regions given their strong association with $O_3$ production [31-33].

The OH radicals oxidize VOCs to produce peroxy radicals such as $HO_2$, which covert NO to $NO_2$ without consuming $O_3$. Then the $NO_2$ produces $O_3$ through photolysis reactions, leading the $O_3$ accumulation [34, 35].

At the same time, the $NO_3$ radical, the primary nighttime oxidant, saw significantly increased levels in the NCP and YRD regions during the Lockdown, with respective growth rates of 0.49 and 0.29 ppt due to the relatively low levels of VOC and $NO_2$, both of which serve as important sinks for the $NO_3$ radical [36] (Fig. S6a). In contrast, in the PRD region, levels of the $NO_3$ radical declined (up to -0.21 ppt) (Fig. 3b-d). Given the enhanced AOC, a significant increase in $O_3$ was observed in the NCP and YRD regions. In the PRD, however, the constant AOC was responsible for a slight decrease in $O_3$. Importantly, in the NCP and YRD regions, the increase in $O_3$ enhanced the AOC due to local photochemistry [37, 38], creating a vicious cycle of heightening $O_3$ levels.

**$O_3$ control policy implications.** Based on this analysis, we have devised a conceptual scheme to demonstrate the roles of the $O_3$ formation regime shift and enhanced AOC on $O_3$ level during the COVID-19 Lockdown period in China (Fig. 4). Satellite data and model simulations have revealed a pronounced shift in the $O_3$ formation regime during the 2020 lockdown period from VOC-limited to $NO_x$-limited, due to the significant decline in $NO_2$. Simultaneously, throughout much of China, the AOC was enhanced, facilitating the secondary formation of $O_3$ and HCHO in the atmosphere. Given the synergistic effects of enhanced AOC and the $O_3$ formation regime shift, a substantial increase in $O_3$ has occurred during the 2020 Lockdown period in much of China, particularly in the NCP and YRD regions. In contrast, $O_3$ levels during this period remained relatively stable in the PRD region. Our results have emphasized the importance of balanced emission control policies for reducing the $O_3$ pollution events in China. Previous policies, which have focused on the arbitrary reduction of primary emission of $NO_x$, $SO_2$, and VOCs, need to be reconsidered as different regions have different $O_3$ sensitivity, and current policies might unintentionally enhance AOC in certain regions, thereby heightening ozone levels. With the synergistic effects of $O_3$ formation regime and AOC, $O_3$ might be elevated in turn. In the future, we recommend that $O_3$ control policies of emission reduction utilize knowledge regarding the synergistic effect of $O_3$ formation regime and AOC variations in the atmosphere on $O_3$ levels. Specifically, we believe emission control policies ought to ensure a balance between emitted $NO_x$ and VOCs to maintain stable $O_3$ formation regimes and thereby control $O_3$ emissions.

## Figures

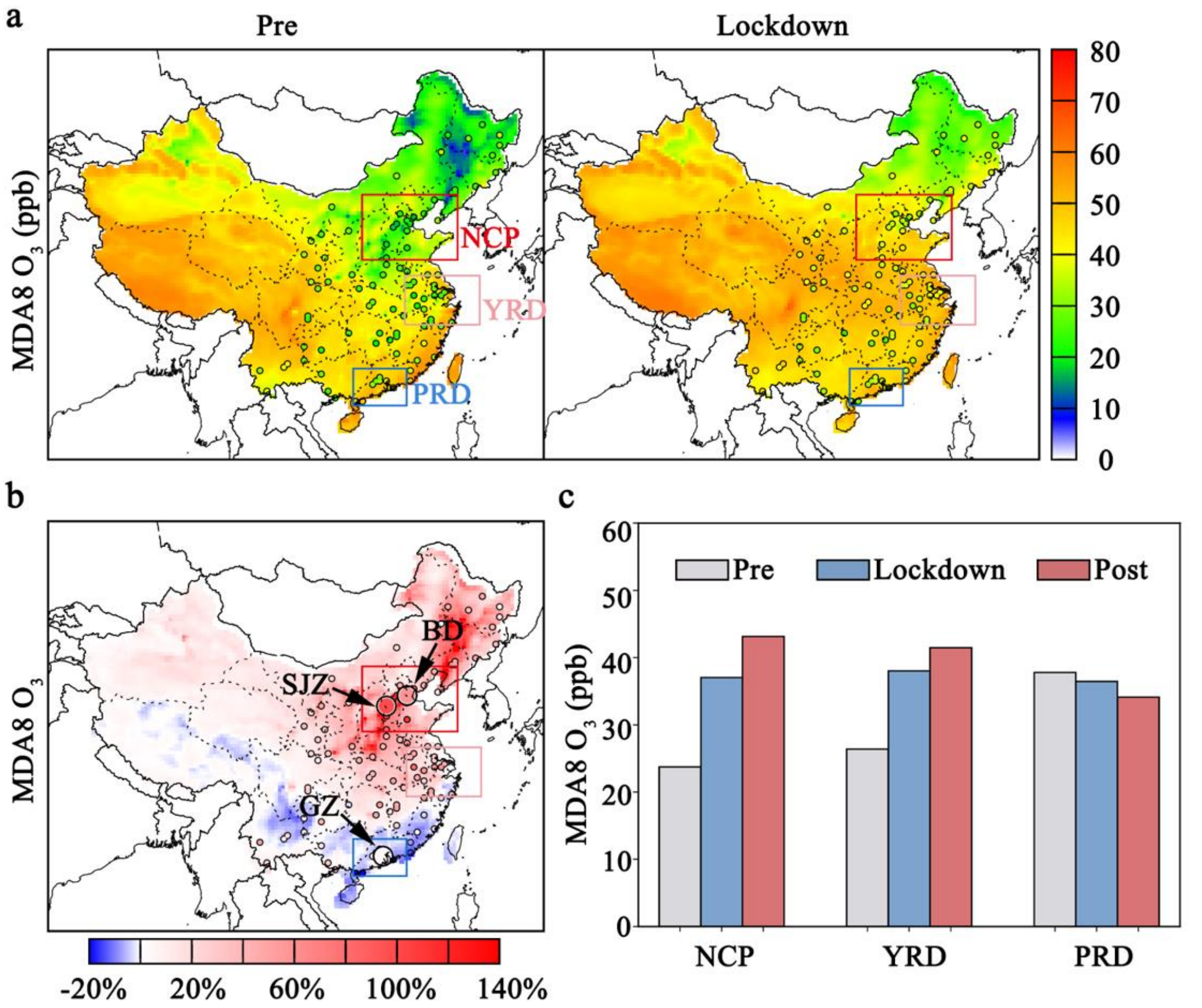

**Fig. 1 MDA8 $O_3$ changes in China during the Pre-lockdown and Lockdowns. (a) CMAQ** predicted and observed surface MDA8 $O_3$ in China during Pre-lockdown and Lockdown periods. The dots represent the observed MDA8 $O_3$ values; SJZ: Shijiazhuang; BD: Baoding; GZ: Guangzhou. **(b)** Observed and simulated MDA8 $O_3$ growth rate during Pre-lockdown and Lockdown periods. **(c)** Observed mean MDA8 $O_3$ in the NCP, YRD, and PRD regions.

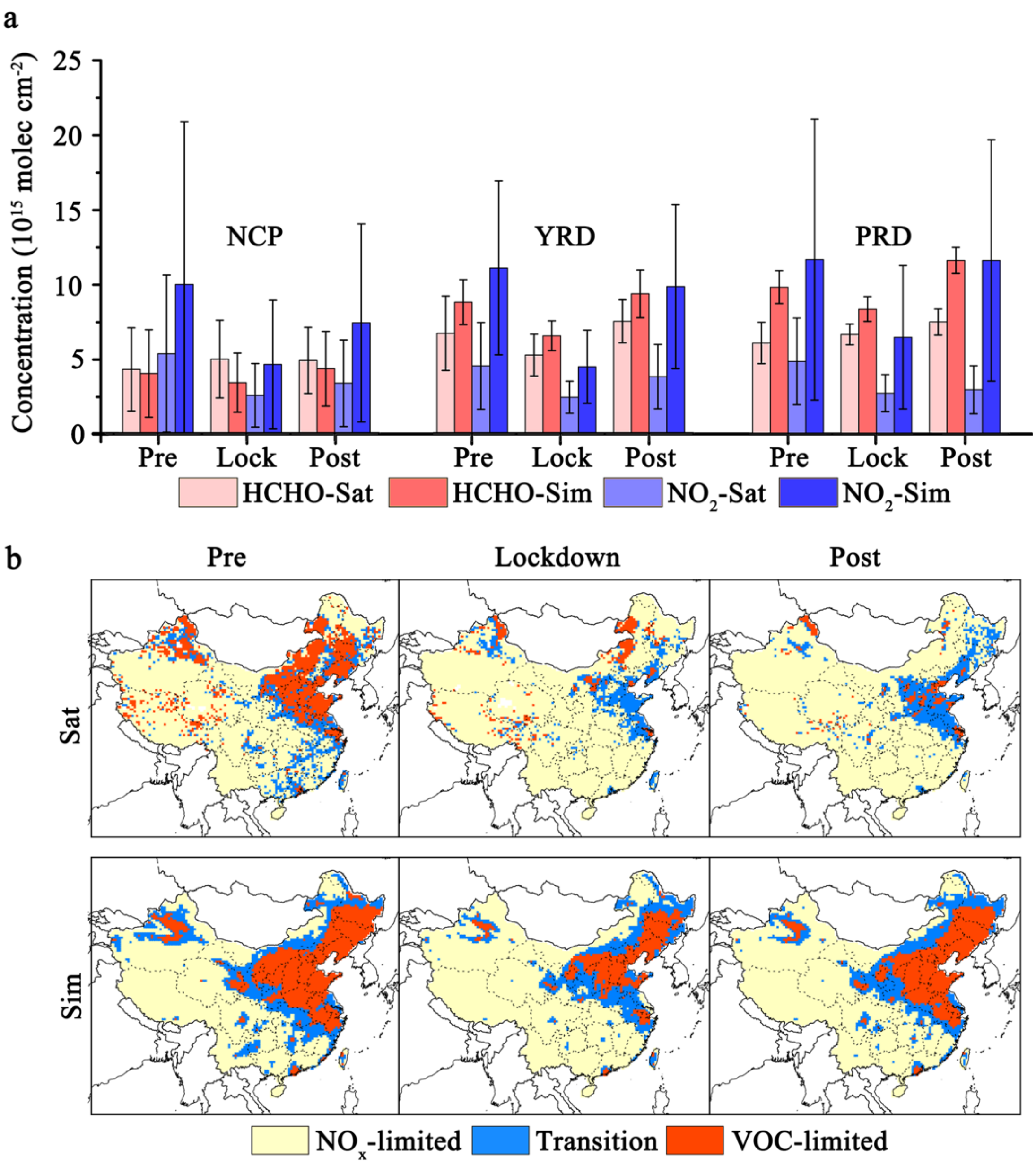

**Fig. 2 The spatial distribution of $NO_2$ and HCHO and the $O_3$ formation regime in China. (a)** The CMAQ predicted results and Tropomi satellite results for $NO_2$ and HCHO column concentrations in NCP, YRD, and PRD regions during the Pre-lockdown, Lockdown, and Post-lockdown periods. **(b)** The predicted and satellite observed results of the $O_3$ formation regime in the troposphere during the Pre-lockdown, Lockdown, and Post-lockdown periods in China.

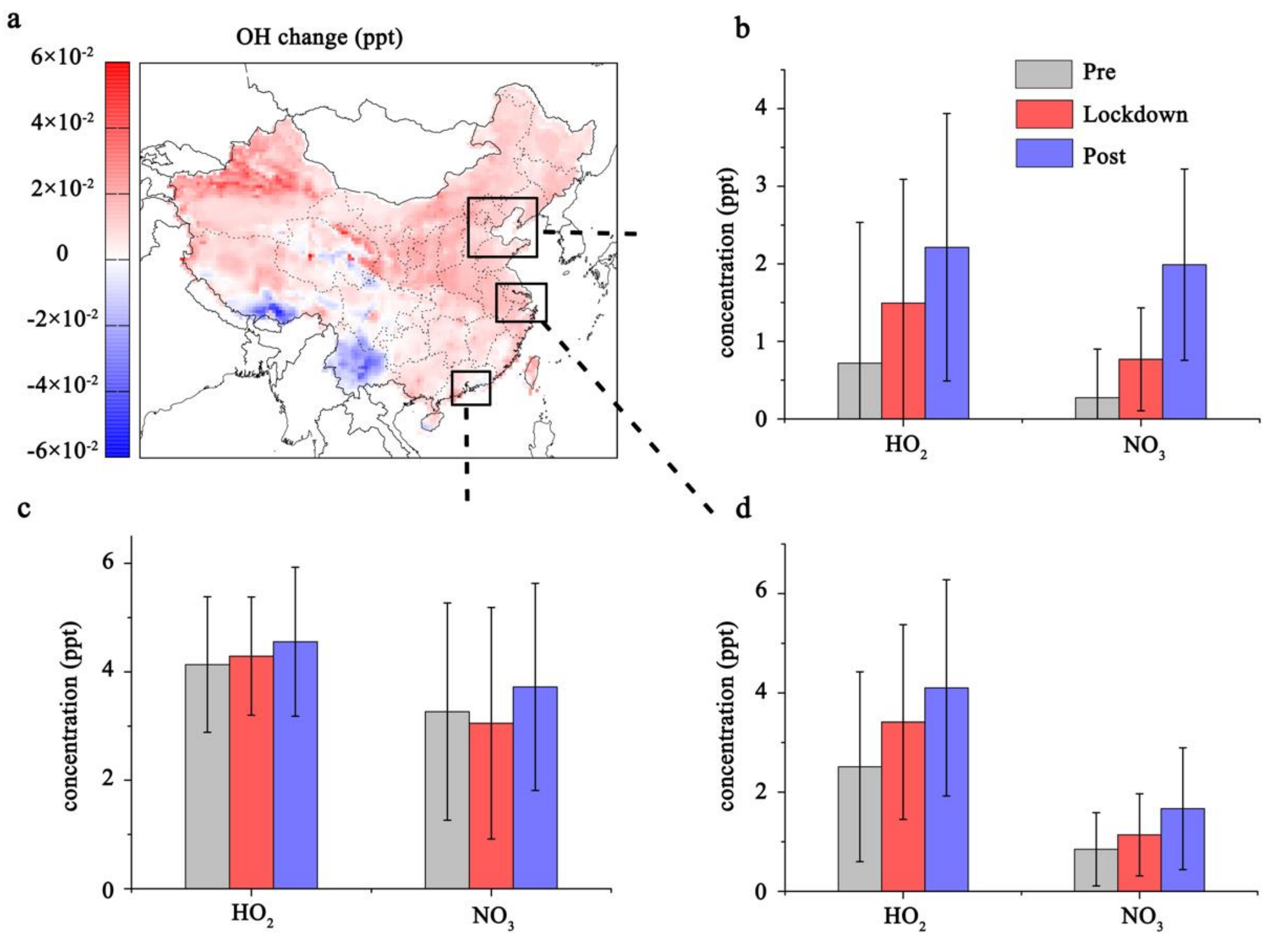

**Fig. 3 Major oxidants changes during the Pre-lockdown and Lockdown. (a)** The spatial distribution variation of simulated OH radical during the Lockdown and Pre-Lockdown periods. **(b-d)** The averaged ground-level concentrations of $HO_2$ and $NO_3$ radicals during the Pre-lockdown, Lockdown, and Post-lockdown periods in NCP, YRD, and PRD regions.

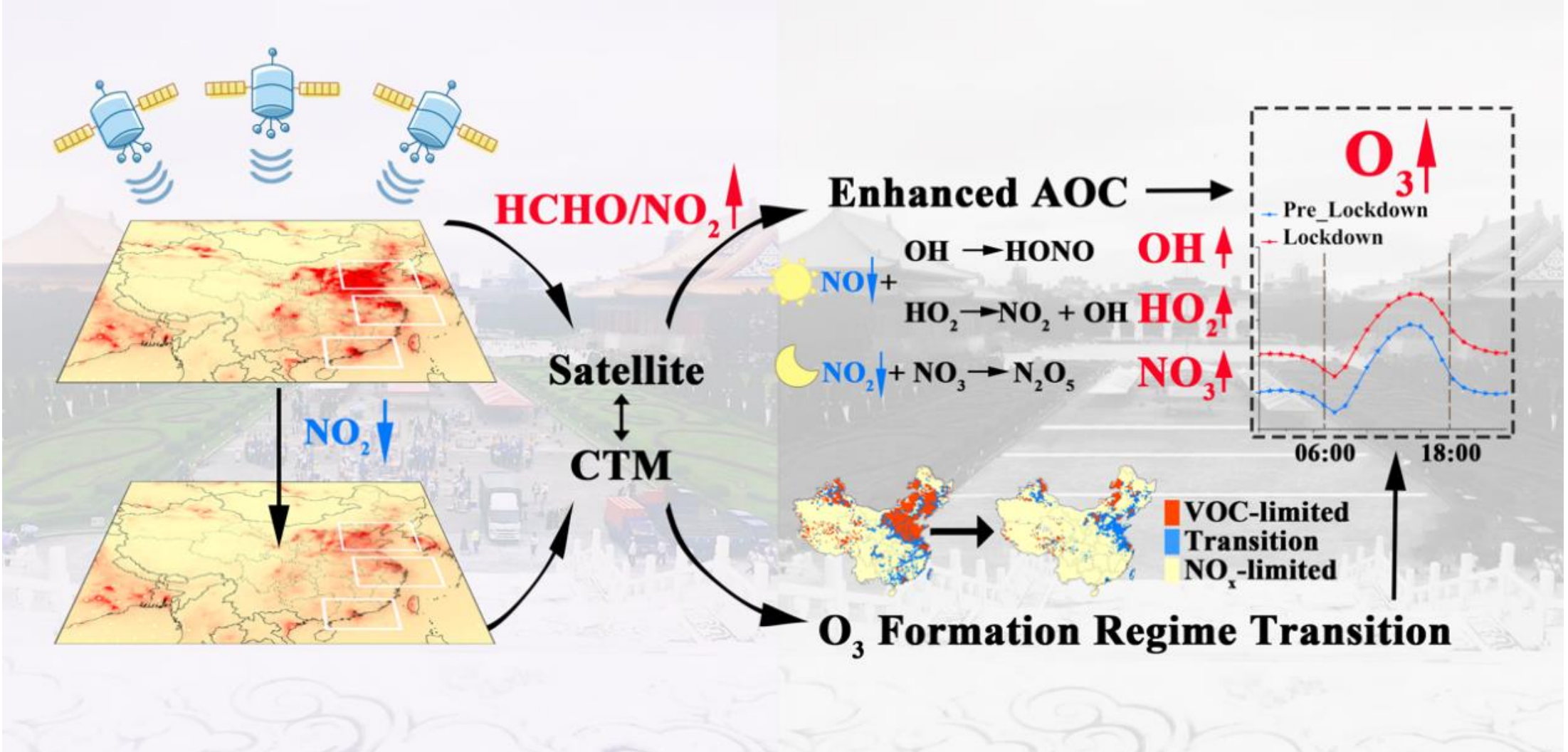

**Fig. 4 Conceptual frame representing the synergistic effect of enhanced AOC and $O_3$ formation regime shift.** The left panel illustrates the decline in $NO_2$ during the Lockdown. The right panel demonstrates the synergistic effect of enhanced AOC and the $O_3$ formation regime shift responsible for the increase in $O_3$.

**Supplementary Information for**

# The "seesaw" impacts between reduced emissions and enhanced AOC on $O_3$ during the COVID-19

Shengqiang Zhu[1], James Poetzscher[1,2], Juanyong Shen[3], Siyu Wang[1], Peng Wang[4], Hongliang Zhang[1,5]

[1]Department of Environmental Science and Engineering, Fudan University, Shanghai 200438, China

[2] School of Environmental Science and Engineering, Nanjing University of Information Science and Technology, 219 Ningliu Road, Nanjing 210044, China

[3]School of Environmental Science and Engineering, Shanghai Jiao Tong University, Shanghai 200240, China

[4]Department of Civil and Environmental Engineering, Hong Kong Polytechnic University, Hong Kong 99907, China

[5]Institute of Eco-Chongming (IEC), Shanghai 200062, China

*Corresponding authors: Peng Wang, peng.ce.wang@polyu.edu.hk; Hongliang Zhang, zhanghl@fudan.edu.cn.

This file includes Supplementary text, and Supplementary Figure S1-S6, Supplementary Table S1-S7

## Supplementary text

### Method

The Sentinel-5 Precursor (S-5P) is a satellite launched by the European Space Agency (ESA) in October 2017 to monitor the spatiotemporal variation of air pollutants, greenhouse gases, UV radiation, and clouds, worldwide. The onboard sensor, TROPOMI (TROPOspheric Monitoring Instrument), has a sun-synchronous orbit with a local overpass time of approximately 13:30 and near-daily global coverage[1-3]. TROPOMI data products are significantly improved compared to the still operating OMI sensor, launched in 2004; specifically, TROPOMI data products feature significantly increased spatial resolution and 1~5 times higher signal-to-noise ratios, compared to OMI products[4]. On August 6, 2019, TROPOMI began outputting data at a higher resolution than before (each near-nadir pixel size is now roughly 5.5 × 3.5 $km^2$)[5].

TROPOMI products are available for free through the Copernicus Open Access Hub (https://scihub.copernicus.eu, last access: 30th July, 2020). Google Earth Engine (https://developers.google.com/earth-engine/datasets/catalog, last access: 30th July, 2020) has converted Level 2 TROPOMI data, provided by the Copernicus Open Access Hub, into the level 3 data used in this study through the bin_spatial operation for the harpconvert tool. In this study, we utilized two TROPOMI datasets from Google Earth Engine, tropospheric NO2 column number density, and tropospheric HCHO column number density, and re-projected them into the same domain as model simulations by using a Lambert projection As suggested by Copernicus, Google Earth Engine filters the source data to remove pixels with QA values less than 75% for tropospheric NO2 column number density datasets and 50% for tropospheric HCHO column number density.

### Model configuration and validation

A modified CMAQ model v5.0.2 with an expanded SAPRC-99 photochemical mechanism was applied to simulate the $O_3$ levels and track the sources of its precursors in China. The details of the source-tracking technique have been described in many previous studies[6-9] and thus are not discussed here. The time interval for which the simulation was conducted spanned from January 1 to March 31, comprising the Pre-lockdown (January 6 to 22), Lockdown (January 23 to February 29), and Post-lockdown (March 1 to 31) periods. Simulations for the same period in 2019, a control period during which there were no emission reductions due to the COVID-19 induced lockdowns, were also conducted. The first five days (January 1-5 in 2019 and 2020) during the simulation periods were chosen as spin-up and removed from the subsequent analysis. The model domain included China and its surrounding countries (Fig. S1), with a horizontal resolution of 36 km × 36 km (127 × 197 grids). The vertical extent was ~20 km from the surface and divided into 18 sigma layers with the first layer height at a height of ~35 m from the surface. The anthropogenic emissions comprised of agriculture, industry, power, residential, and transportation sectors were based on the Multi-resolution Emission Inventory for China (MEIC, v1.3, 0.25° × 0.25°, http://www.meicmodel.org). The anthropogenic emissions of different pollutants in China during the Lockdown were adjusted according to a recent study to reflect the reduced human activity (Table S1)[10]. Emissions from other countries were obtained from the Emissions Database for Global Atmospheric Research (EDGAR) v4.3.1 (https://edgar.jrc.ec.europa.eu/overview.php?v=431). Biogenic emissions were provided by the Model of Emissions of Gases and aerosols from Nature (MEGAN) v2.1[11]. The meteorological inputs were generated by the Weather Research and Forecasting (WRF) model v4.2 (https://www2.mmm.ucar.edu/wrf/users), with the initial and boundary conditions based on the National Centers for Environmental Prediction (NCEP) Final (FNL) Operational Model Global Tropospheric Analyses dataset (https://rda.ucar.edu/datasets/ds083.2).

The WRF model performance was evaluated by comparing the predicted temperature (T), relative humidity (RH) at 2 m above the surface, wind speed (WS), and wind direction (WD) at 10 m above the surface with observation data from the National Climatic Data Center (ftp://ftp.ncdc.noaa.gov/pub/data/noaa/) (~1400 sites) (Table S2, S4). The model slightly underestimated the T during the simulation periods, with the mean bias (MB) values of -1.35 to -0.43 similar to a previous study[12]. The gross error (GE) values (1.65-1.93) of WS were all within the benchmark[13]. The WD was generally predicted successfully, with values around the benchmarks. The simulation of RH was comparable with previous studies[12, 14, 15]. Overall, the WRF model performed well, providing the CMAQ model with reliable meteorological inputs.

The CMAQ model simulations were validated using the observations (~1500 sites) from the national air quality monitoring network (http://www.cnemc.cn/) (Table S3-S5). The model accurately predicted the hourly $O_3$ concentrations, with mean normalized bias (MNB) values of 0.04-0.13 and mean normalized error (MNE) values ranging from 0.09 to 0.17, which were all within the criteria[16, 17]. The simulations of $O_3$-8h were slightly overestimated with MNB values of 0.18-0.21, a little higher than the criteria.

The changes in $O_3$, $NO_2$, and $SO_2$ during the simulation periods were accurately captured by the model. In general, the model performed well, accurately simulating $O_3$ levels throughout the analyzed time periods.

$NO_2$ and HCHO concentrations from 17 vertical layers (with the highest layer having a height of ~10km) in the CMAQ model were added up to ascertain their tropospheric column concentrations. The $NO_2$ and HCHO column concentrations generated by the model were then compared with the satellite data to determine the shift in $O_3$ formation regimes. In our study, the $O_3$ formation regimes were categorized into VOC-limited, $NO_x$-limited and transition regimes based on the formaldehyde nitrogen ratio (FNR)[18, 19]. Here we set FNR $< 1.0$ as a VOC-limited regime, FNR $> 2.0$ as a $NO_x$-limited regime and FNR between 1.0 and 2.0 as a transitional regime[20, 21].

**Supplement formula**

The mean bias (MB), gross error (GE), and root mean square error (RMSE) were used for the evaluation of WRF model performance, while the mean normalized bias (MNB), mean normalized error (MNE), mean fractional bias (MFB) and mean fractional error (MFE) were used for the validation of CMAQ model performance.

$$MB = \frac{1}{N}\sum_{i=1}^{N}\left(C_{m,i} - C_{o,i}\right) \tag{1}$$

$$GE = \frac{1}{N}\sum_{i=1}^{N}\left|C_{m,i} - C_{o,i}\right| \tag{2}$$

$$RMSE = \left(\frac{1}{N}\sum_{i=1}^{N}\left(C_{m,i} - C_{o,i}\right)^{2}\right)^{\frac{1}{2}} \tag{3}$$

$$MNB = \frac{1}{N}\sum_{i=1}^{N}\frac{C_{m,i} - C_{o,i}}{C_{o,i}} \tag{4}$$

$$MNE = \frac{1}{N}\sum_{i=1}^{N}\frac{\left|C_{m,i} - C_{o,i}\right|}{C_{o,i}} \tag{5}$$

$$MFB = \frac{2}{N}\sum_{i=1}^{N}\frac{C_{m,i} - C_{o,i}}{C_{m,i} + C_{o,i}} \tag{6}$$

$$MFE = \frac{2}{N}\sum_{i=1}^{N}\frac{\left|C_{m,i} - C_{o,i}\right|}{C_{m,i} + C_{o,i}} \tag{7}$$

where $C_{m,i}$ and $C_{o,i}$ refer to the i$^{th}$ predicted and observed value, respectively. $N$ is the number of prediction-observation pairs drawn from all measurement sites.

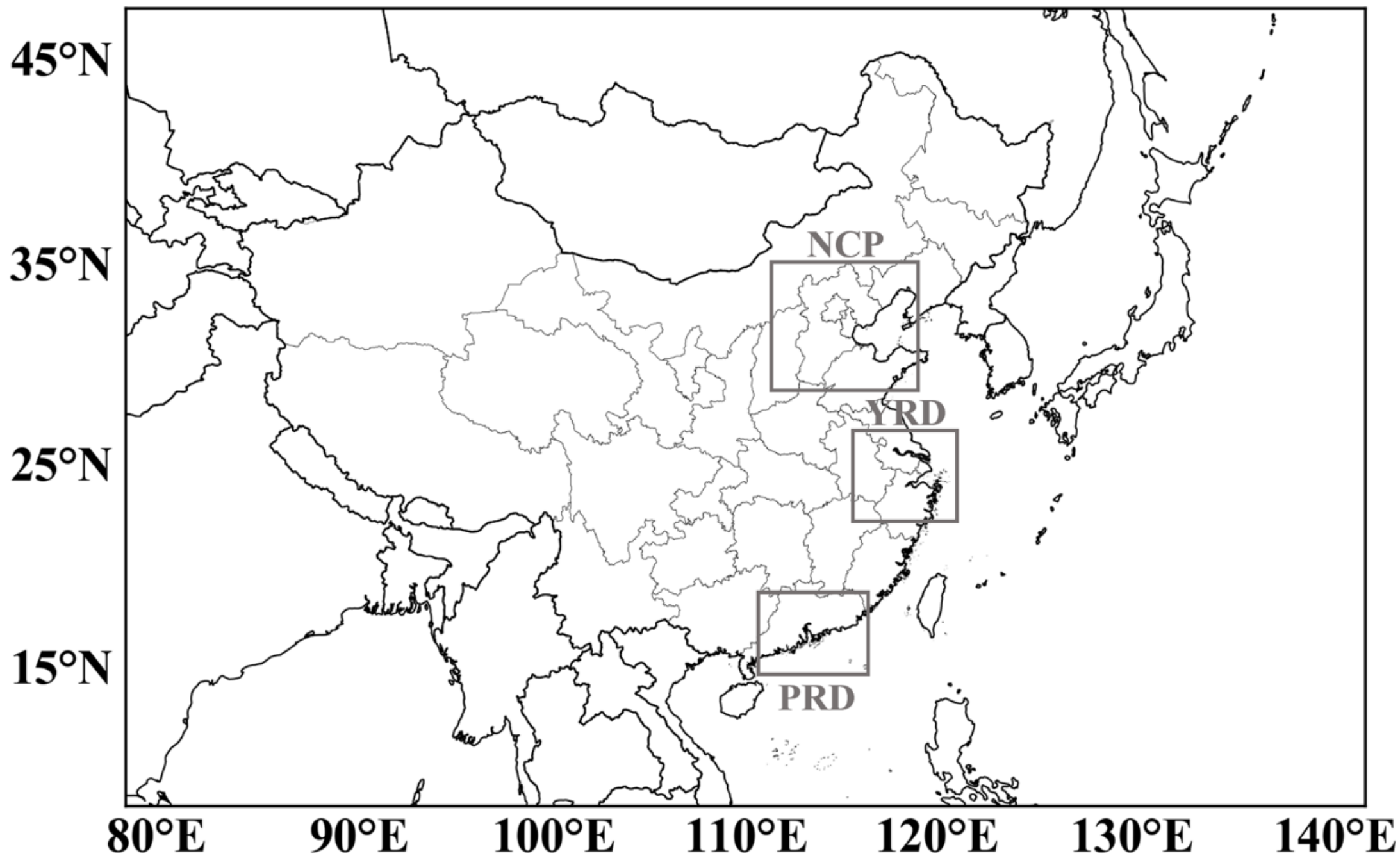

Figure S1: Model domain and three subregions (NCP, YRD, and PRD).

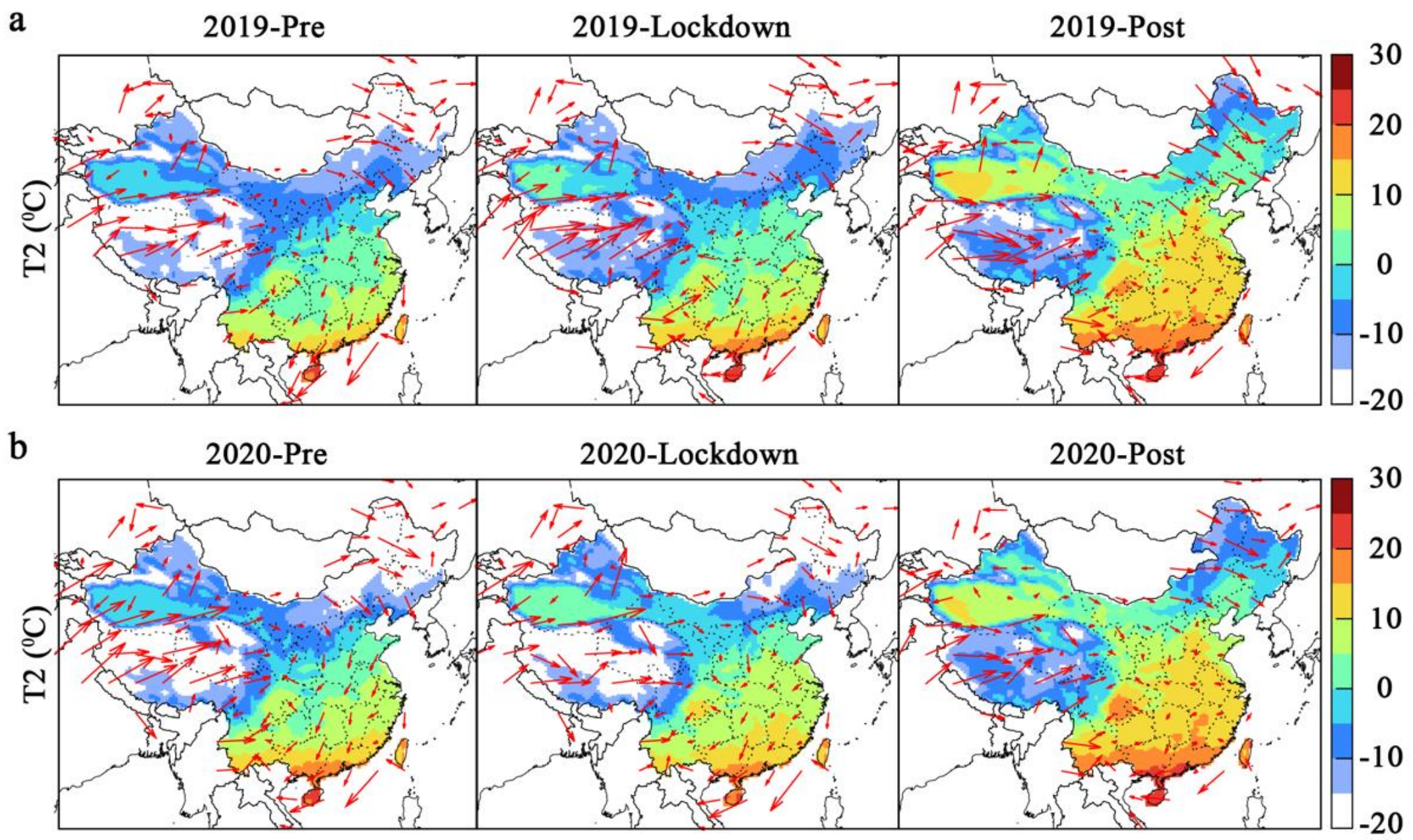

Figure S2: Regional variation of the temperature at 2 meters above ground and wind speed during the Pre-Lockdown, Lockdown, and Post-Lockdown periods in 2020 and the same periods in 2019.

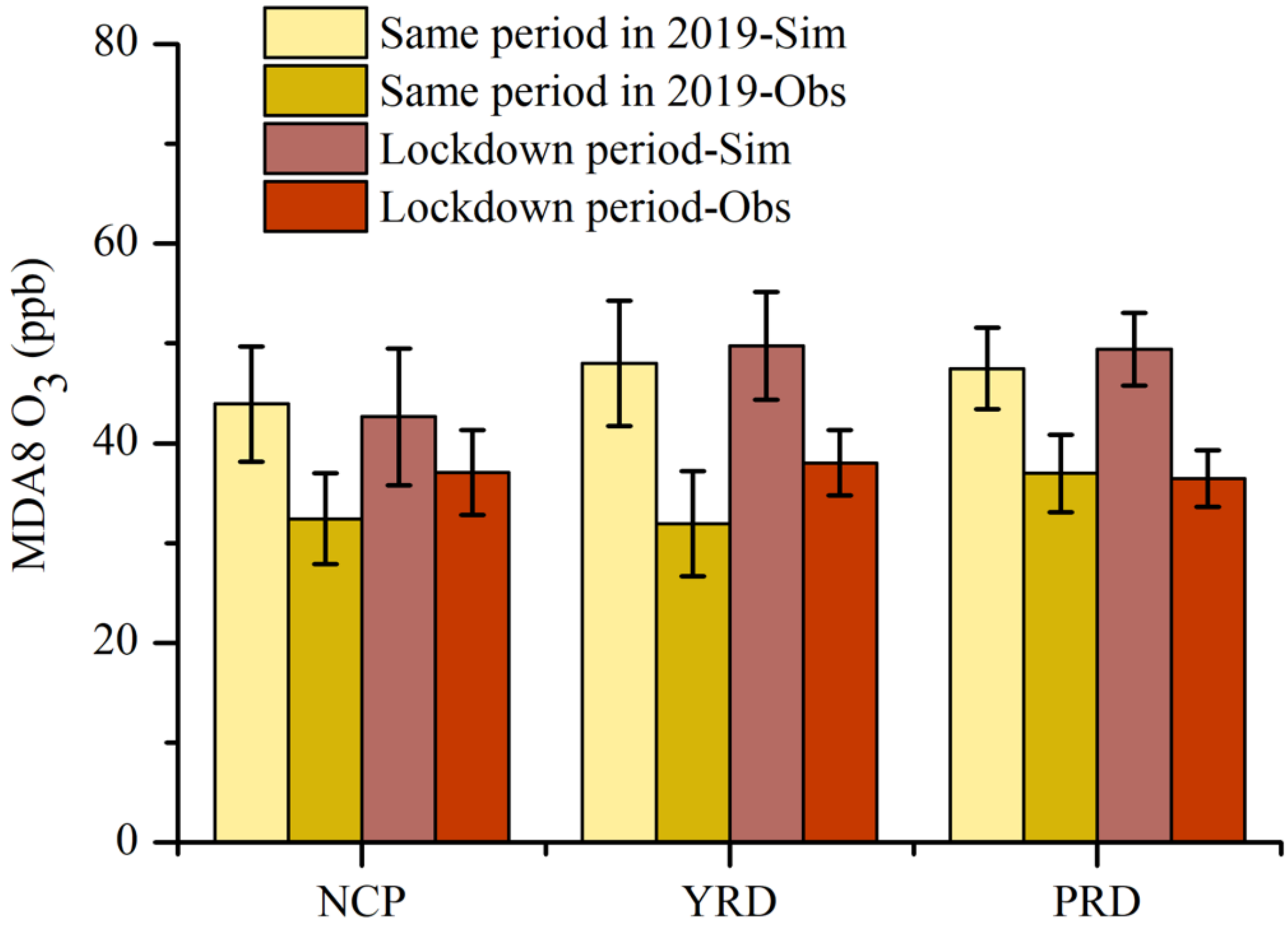

Figure S3: Simulated and observed mean MDA8 $O_3$ in the three key regions during the Lockdown and the same period in 2019.

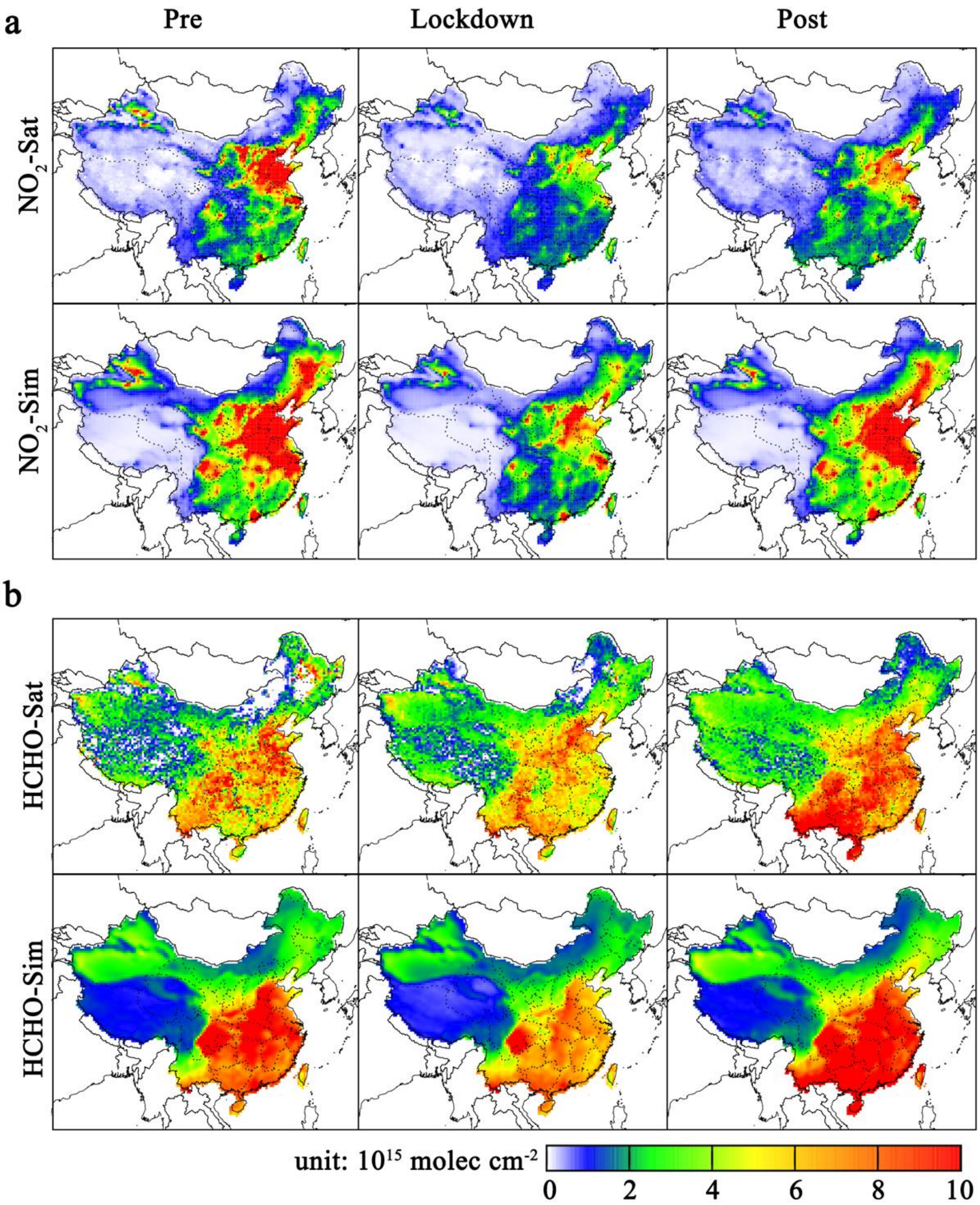

Figure S4: Simulated and satellite observed spatial distribution of $NO_2$ and HCHO during the Pre-Lockdown, Lockdown, and Post-Lockdown periods in 2020.

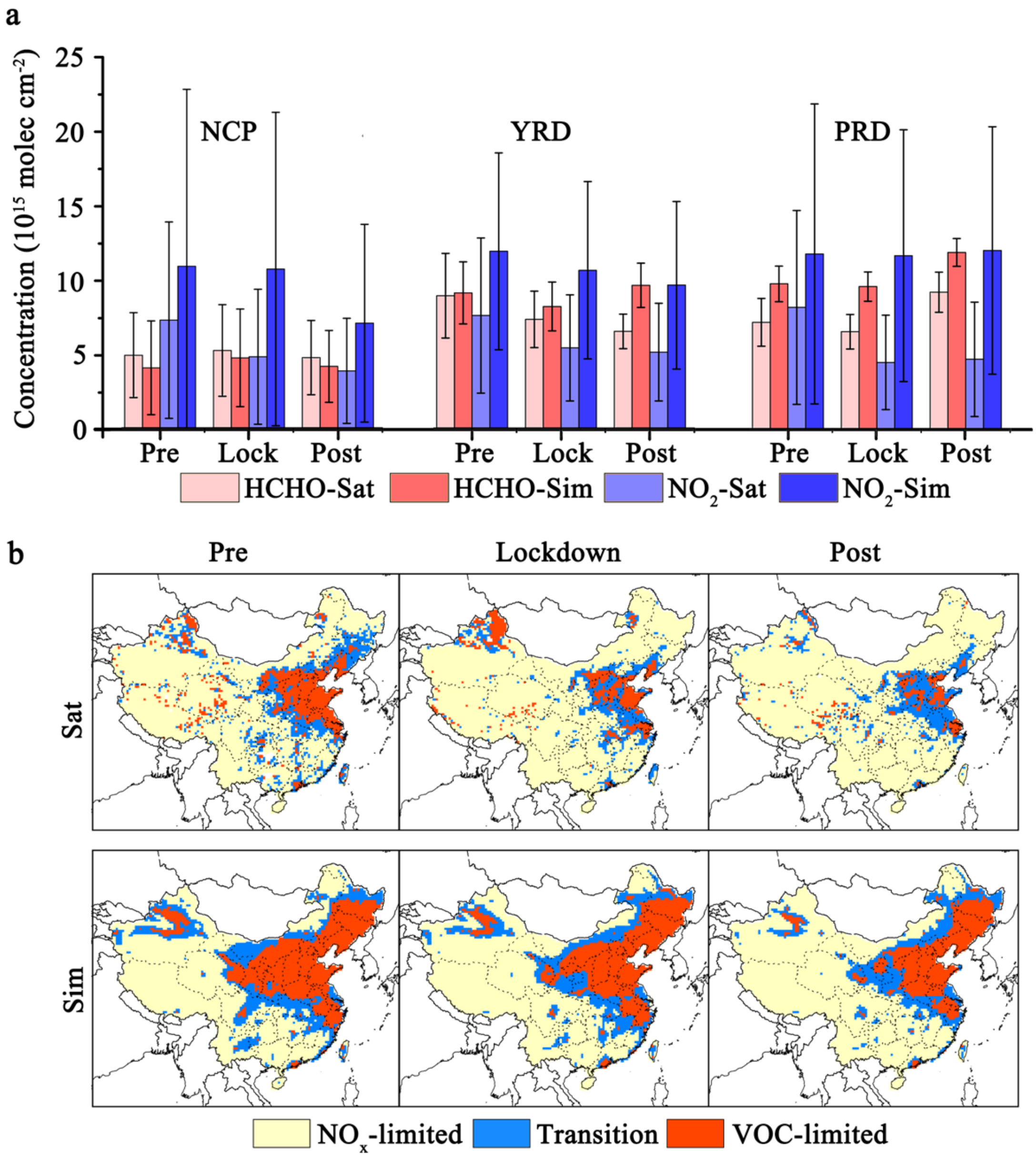

Figure S5: The spatial distribution of $NO_2$ and HCHO and the $O_3$ formation regime in China (a) The simulated and Tropomi satellite observed results of $NO_2$ and HCHO column concentrations in NCP, YRD, and PRD regions during the same periods of Pre-Lockdown, Lockdown, and Post-Lockdown in 2019. (b) The simulated and satellite observed results of $O_3$ formation regime in the troposphere during the same periods of Pre-Lockdown, Lockdown, and Post-Lockdown in China.

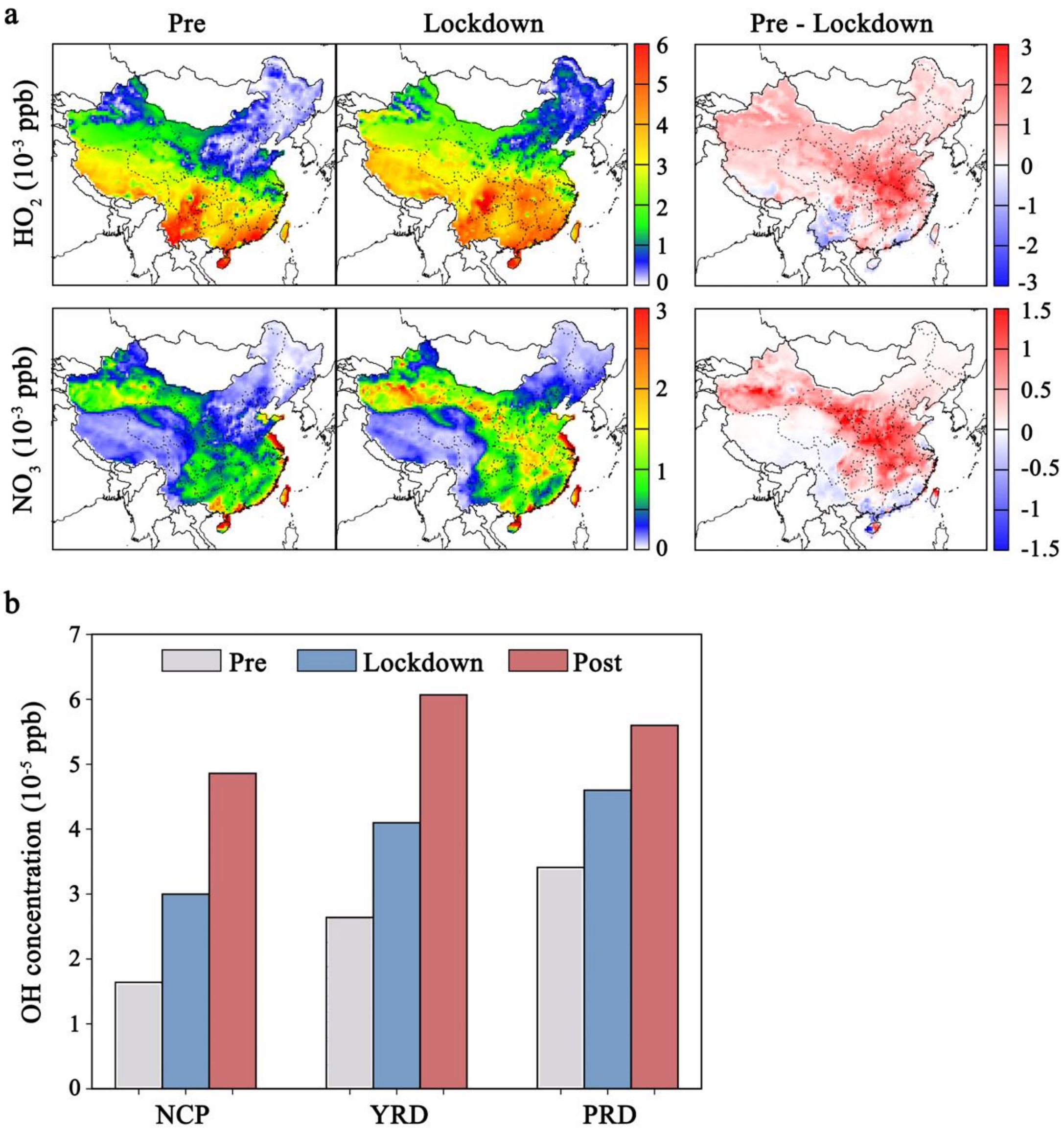

Figure S6: Major oxidants changes during the Pre-lockdown and Lockdown. (a) Regional variation of the $HO_2$ and $NO_3$ radicals during the Pre-Lockdown and Lockdown periods in 2020. (b) The variation in the OH radical concentrations during the Pre-Lockdown, Lockdown, and Post-Lockdown periods in the NCP, YRD, and PRD regions in 2020.

Table S1: Emission reduction ratios of different air pollutants at the provincial level during the Lockdown [10].

| Province | CO | $NO_x$ | $SO_2$ | VOC | $PM_{2.5}$ | BC | OC |
|---|---|---|---|---|---|---|---|
| Beijing | 22% | 45% | 26% | 45% | 18% | 46% | 8% |
| Tianjin | 21% | 38% | 20% | 41% | 14% | 22% | 6% |
| Hebei | 15% | 45% | 16% | 36% | 12% | 17% | 5% |
| Shanxi | 18% | 40% | 20% | 33% | 16% | 19% | 10% |
| Inner Mongolia | 14% | 29% | 15% | 34% | 13% | 16% | 6% |
| Liaoning | 21% | 40% | 28% | 36% | 16% | 28% | 8% |
| Jilin | 16% | 39% | 23% | 34% | 13% | 18% | 5% |
| Heilongjiang | 17% | 37% | 27% | 28% | 13% | 15% | 7% |
| Shanghai | 35% | 48% | 42% | 45% | 34% | 54% | 42% |
| Jiangsu | 23% | 50% | 26% | 41% | 16% | 35% | 7% |
| Zhejiang | 41% | 50% | 29% | 45% | 30% | 49% | 20% |
| Anhui | 14% | 56% | 22% | 31% | 11% | 22% | 4% |
| Fujian | 29% | 51% | 30% | 42% | 19% | 31% | 7% |
| Jiangxi | 24% | 53% | 21% | 43% | 19% | 30% | 9% |
| Shandong | 23% | 50% | 25% | 39% | 19% | 35% | 9% |
| Henan | 23% | 57% | 22% | 41% | 18% | 35% | 8% |
| Hubei | 19% | 55% | 23% | 35% | 16% | 23% | 10% |
| Hunan | 22% | 51% | 25% | 36% | 20% | 24% | 15% |
| Guangdong | 38% | 50% | 33% | 46% | 27% | 42% | 13% |
| Guangxi | 24% | 50% | 28% | 39% | 17% | 27% | 5% |
| Hainan | 24% | 44% | 25% | 36% | 14% | 25% | 4% |
| Chongqing | 18% | 53% | 32% | 37% | 14% | 20% | 4% |
| Sichuan | 16% | 50% | 27% | 33% | 9% | 15% | 3% |
| Guizhou | 24% | 39% | 25% | 30% | 22% | 25% | 20% |
| Yunnan | 24% | 51% | 25% | 41% | 18% | 21% | 8% |
| Tibet | 16% | 35% | 15% | 35% | 14% | 14% | 5% |
| Shaanxi | 19% | 45% | 18% | 34% | 13% | 22% | 5% |
| Gansu | 13% | 47% | 16% | 29% | 9% | 13% | 3% |
| Qinghai | 23% | 46% | 22% | 39% | 20% | 20% | 7% |
| Ningxia | 24% | 36% | 24% | 39% | 20% | 23% | 8% |
| Xinjiang | 16% | 35% | 15% | 35% | 14% | 14% | 5% |

Table S2: Model performance of meteorological parameters temperature (T), wind speed (WS), wind direction (WD), and relative humidity (RH) from January to March in 2020 (PRE is mean prediction; OBS is mean observation; MB is mean bias; GE is gross error; and RMSE is root mean square error).

| | | Jan | Feb | Mar | Benchmark* |
|---|---|---|---|---|---|
| T (K) | PRE | 276.55 | 278.15 | 282.79 | |
| | OBS | 277.54 | 279.47 | 284.22 | |
| | MB | **-0.93** | **-1.25** | **-1.35** | $\leq \pm 0.5$ |
| | GE | **2.78** | **2.89** | **2.80** | $\leq 2.0$ |
| | RMSE | 3.79 | 3.93 | 3.84 | |
| WS (m/s) | PRE | 4.55 | 4.66 | 4.68 | |
| | OBS | 3.24 | 3.38 | 3.51 | |
| | MB | **1.32** | **1.28** | **1.17** | $\leq \pm 0.5$ |
| | GE | 1.89 | 1.90 | 1.83 | $\leq 2.0$ |
| | RMSE | **2.45** | **2.45** | **2.35** | $\leq 2.0$ |
| WD (°) | PRE | 191.94 | 197.16 | 197.74 | |
| | OBS | 188.06 | 186.69 | 189.12 | |
| | MB | 8.17 | **11.30** | **10.06** | $\leq \pm 10$ |
| | GE | **45.67** | **45.08** | **44.55** | $\leq 30$ |
| | RMSE | **62.71** | **62.18** | **61.87** | $\leq 30$ |
| RH (%) | PRE | 72.27 | 69.27 | 67.48 | |
| | OBS | 74.24 | 70.96 | 70.08 | |
| | MB | -1.97 | -1.69 | -2.60 | |
| | GE | 11.12 | 11.63 | 12.30 | |
| | RMSE | 14.41 | 15.19 | 15.98 | |

Note: * are benchmarks limits suggested by Emery et al. (2001), data which do not fall under the limits are shown as bold.

Table S3: Model performance on pollutants concentration in China from January to March 2020. (MNB: mean normalized bias; MNE: mean normalized error; MFB: mean fractional bias; and MFE: mean fractional error).

| | | Jan | Feb | Mar | Criteria |
|---|---|---|---|---|---|
| | OBS | 64.68 | 64.95 | 67.10 | |
| | PRE | 69.54 | 67.28 | 75.52 | |
| $O_3$-1h | MNB | 0.08 | 0.04 | 0.13 | ≤±0.15 |
| (ppb) | MNE | 0.12 | 0.09 | 0.17 | ≤0.3 |
| | MFB | 0.07 | 0.03 | 0.11 | |
| | MFE | 0.11 | 0.09 | 0.15 | |
| | OBS | 42.56 | 43.76 | 46.61 | |
| | PRE | 49.89 | 50.92 | 55.75 | |
| $O_3$-8h | MNB | **0.19** | **0.18** | **0.21** | ≤±0.15 |
| (ppb) | MNE | 0.24 | 0.21 | 0.24 | ≤0.3 |
| | MFB | 0.15 | 0.15 | 0.17 | |
| | MFE | 0.20 | 0.19 | 0.21 | |
| | OBS | 14.98 | 9.13 | 12.26 | |
| | PRE | 10.43 | 6.66 | 10.50 | |
| $NO_2$ | MNB | -0.15 | -0.08 | 0.03 | |
| (ppb) | MNE | 0.64 | 0.73 | 0.73 | |
| | MFB | -0.48 | -0.48 | -0.36 | |
| | MFE | 0.73 | 0.80 | 0.73 | |
| | OBS | 4.53 | 3.61 | 3.46 | |
| | PRE | 8.83 | 5.88 | 6.47 | |
| $SO_2$ | MNB | 2.03 | 1.47 | 1.66 | |
| (ppb) | MNE | 2.38 | 1.91 | 2.05 | |
| | MFB | 0.32 | 0.15 | 0.22 | |
| | MFE | 0.86 | 0.85 | 0.84 | |

Table S4: Model performance of meteorological parameters temperature (T), wind speed (WS), wind direction (WD), and relative humidity (RH) from January to March in 2019 (PRE is mean prediction; OBS is mean observation; MB is mean bias; GE is gross error; and RMSE is root mean square error).

| | | Jan | Feb | Mar | Benchmark* |
|---|---|---|---|---|---|
| T (K) | PRE | 276.48 | 278.04 | 283.04 | |
| | OBS | 276.99 | 278.74 | 283.91 | |
| | MB | -0.43 | **-0.61** | **-0.78** | ⩽±0.5 |
| | GE | **2.64** | **2.65** | **2.55** | ⩽2.0 |
| | RMSE | 3.64 | 3.61 | 3.52 | |
| WS (m/s) | PRE | 4.77 | 4.52 | 4.64 | |
| | OBS | 3.46 | 3.39 | 3.56 | |
| | MB | **1.32** | **1.13** | **1.08** | ⩽±0.5 |
| | GE | 1.93 | 1.80 | 1.77 | ⩽2.0 |
| | RMSE | **2.50** | **2.35** | **2.30** | ⩽2.0 |
| WD (°) | PRE | 208.85 | 196.46 | 205.49 | |
| | OBS | 198.21 | 189.31 | 194.23 | |
| | MB | **12.52** | 9.68 | **11.61** | ⩽±10 |
| | GE | **44.53** | **45.57** | **44.30** | ⩽30 |
| | RMSE | **61.42** | **62.56** | **61.33** | ⩽30 |
| RH (%) | PRE | 67.75 | 68.41 | 64.31 | |
| | OBS | 72.39 | 72.88 | 68.12 | |
| | MB | -4.64 | -4.47 | -3.81 | |
| | GE | 11.92 | 11.88 | 12.24 | |
| | RMSE | 15.92 | 15.46 | 15.80 | |

Note: * are benchmarks limits suggested by Emery et al. (2001), data which do not fall under the limits are shown as bold.

Table S5: Model performance on pollutants concentration in China from January to March 2019. (MNB: mean normalized bias; MNE: mean normalized error; MFB: mean fractional bias; and MFE: mean fractional error).

| | | Jan | Feb | Mar | Criteria |
|---|---|---|---|---|---|
| | OBS | 66.51 | 67.69 | 67.03 | |
| | PRE | 72.86 | 74.08 | 79.14 | |
| $O_3$-1h | MNB | 0.11 | 0.11 | **0.19** | ≤±0.15 |
| (ppb) | MNE | 0.15 | 0.16 | 0.21 | ≤0.3 |
| | MFB | 0.09 | 0.09 | 0.16 | |
| | MFE | 0.14 | 0.15 | 0.18 | |
| | OBS | 42.66 | 42.95 | 48.14 | |
| | PRE | 52.57 | 52.03 | 60.37 | |
| $O_3$-8h | MNB | **0.25** | **0.23** | **0.27** | ≤±0.15 |
| (ppb) | MNE | 0.27 | 0.28 | 0.30 | ≤0.3 |
| | MFB | 0.20 | 0.18 | 0.21 | |
| | MFE | 0.23 | 0.23 | 0.24 | |
| | OBS | 19.77 | 13.00 | 15.84 | |
| | PRE | 12.88 | 11.55 | 11.24 | |
| $NO_2$ | MNB | -0.23 | 0.18 | -0.15 | |
| (ppb) | MNE | 0.61 | 0.83 | 0.69 | |
| | MFB | -0.55 | -0.27 | -0.52 | |
| | MFE | 0.75 | 0.72 | 0.79 | |
| | OBS | 5.98 | 4.54 | 4.16 | |
| | PRE | 10.56 | 7.27 | 7.30 | |
| $SO_2$ | MNB | 1.92 | 1.55 | 1.58 | |
| (ppb) | MNE | 2.25 | 1.96 | 1.99 | |
| | MFB | 0.32 | 0.19 | 0.18 | |
| | MFE | 0.84 | 0.83 | 0.83 | |

Table S6: Source apportionment analysis of $NO_2$ and HCHO during the Pre-Lockdown, Lockdown, and Post-Lockdown periods in 2020. Units are $10^{15}$ molec/cm$^2$.

| Pre-Lockdown | Background | Power | Agriculture | Industry | Residential | Transportation | Others | Total |
|---|---|---|---|---|---|---|---|---|
| $NO_2$(NCP) | 0.08 | 2.77 | 0.0 | 3.97 | 0.74 | 2.68 | 0.08 | 10.32 |
| $NO_2$(YRD) | 0.08 | 2.80 | 0.0 | 4.18 | 0.57 | 4.10 | 0.23 | 11.96 |
| $NO_2$(PRD) | 0.11 | 1.90 | 0.0 | 2.78 | 0.14 | 4.09 | 2.09 | 11.11 |
| HCHO(NCP) | 2.57 | 0.0 | 0.0 | 0.52 | 0.75 | 0.28 | 0.12 | 4.24 |
| HCHO(YRD) | 5.24 | 0.0 | 0.0 | 1.09 | 1.46 | 0.59 | 0.48 | 8.86 |
| HCHO(PRD) | 6.96 | 0.0 | 0.0 | 0.82 | 0.43 | 0.53 | 0.89 | 9.63 |
| Lockdown | Background | Power | Agriculture | Industry | Residential | Transportation | Others | Total |
| $NO_2$(NCP) | 0.09 | 1.15 | 0.0 | 1.82 | 0.32 | 1.27 | 0.06 | 4.71 |
| $NO_2$(YRD) | 0.07 | 0.95 | 0.0 | 1.56 | 0.21 | 1.71 | 0.24 | 4.74 |
| $NO_2$(PRD) | 0.10 | 0.76 | 0.0 | 1.18 | 0.07 | 2.02 | 2.13 | 6.26 |
| HCHO(NCP) | 2.44 | 0.0 | 0.0 | 0.31 | 0.42 | 0.17 | 0.08 | 3.41 |
| HCHO(YRD) | 4.70 | 0.0 | 0.0 | 0.51 | 0.64 | 0.31 | 0.46 | 6.62 |
| HCHO(PRD) | 6.50 | 0.0 | 0.0 | 0.33 | 0.17 | 0.26 | 0.99 | 8.25 |
| Post-Lockdown | Background | Power | Agriculture | Industry | Residential | Transportation | Others | Total |
| $NO_2$(NCP) | 0.08 | 1.83 | 0.0 | 3.01 | 0.33 | 2.03 | 0.07 | 7.35 |
| $NO_2$(YRD) | 0.07 | 2.39 | 0.0 | 3.83 | 0.26 | 3.56 | 0.28 | 10.39 |
| $NO_2$(PRD) | 0.09 | 1.85 | 0.0 | 2.73 | 0.10 | 3.63 | 2.63 | 11.03 |
| HCHO(NCP) | 3.40 | 0.0 | 0.0 | 0.36 | 0.27 | 0.18 | 0.10 | 4.31 |
| HCHO(YRD) | 7.21 | 0.0 | 0.0 | 0.76 | 0.46 | 0.40 | 0.65 | 9.48 |
| HCHO(PRD) | 9.24 | 0.0 | 0.0 | 0.49 | 0.12 | 0.32 | 1.31 | 11.48 |

Table S7: The percentage of each $O_3$ formation regime type in the NCP, YRD and PRD regions during the Pre-Lockdown, Lockdown, and Post-Lockdown periods in 2020.

| Pre-Lockdown | Tro | | | Sim | | |
|---|---|---|---|---|---|---|
| | $NO_x$-limited | Transition | VOC-limited | $NO_x$-limited | Transition | VOC-limited |
| NCP | 0.11 | 0.22 | 0.67 | 0.13 | 0.15 | 0.72 |
| YRD | 0.37 | 0.45 | 0.17 | 0.03 | 0.31 | 0.66 |
| PRD | 0.31 | 0.41 | 0.28 | 0.31 | 0.23 | 0.46 |
| Lockdown | Tro | | | Sim | | |
| | $NO_x$-limited | Transition | VOC-limited | $NO_x$-limited | Transition | VOC-limited |
| NCP | 0.56 | 0.36 | 0.08 | 0.20 | 0.28 | 0.52 |
| YRD | 0.65 | 0.33 | 0.02 | 0.29 | 0.55 | 0.16 |
| PRD | 0.69 | 0.31 | 0.0 | 0.44 | 0.28 | 0.28 |
| Post-Lockdown | Tro | | | Sim | | |
| | $NO_x$-limited | Transition | VOC-limited | $NO_x$-limited | Transition | VOC-limited |
| NCP | 0.48 | 0.36 | 0.16 | 0.09 | 0.26 | 0.65 |
| YRD | 0.57 | 0.38 | 0.05 | 0.09 | 0.39 | 0.51 |
| PRD | 0.74 | 0.0 | 0.26 | 0.28 | 0.36 | 0.36 |